\documentclass{aa}
\usepackage{graphicx}
\begin{document}
   \title{The C star outer disk population of M31 seen with the
SLOAN filters
\thanks{ Based on observations obtained with MegaPrime/MegaCam, 
a joint project of CFHT and CEA/DAPNIA, at the 
Canada-France-Hawaii Telescope (CFHT) which is 
operated by the National Research Council (NRC) of Canada, 
the Institut National des Science de l'Univers of the 
Centre National de la Recherche Scientifique (CNRS) of France, and 
the University of Hawaii.
}}


   \author{S. Demers
          \inst{1},
\and 
          P. Battinelli \inst{2}
          }

   \offprints{S. Demers}

   \institute{D\'epartement de Physique, Universit\'e de Montr\'eal,
                C.P.6128, Succursale Centre-Ville, Montr\'eal,
                Qu\'ebec H3C 3J7, Canada\\
                \email {demers@astro.umontreal.ca }
\and
INAF, Osservatorio Astronomico di Roma,
              Viale del Parco Mellini 84, I-00136 Roma, Italia\\
              \email {battinel@oarhp1.rm.astro.it }
}
   \date{Received; accepted}

   \abstract{ We employ the CFHT Megacam camera to survey $\sim$ one deg$^2$
of the southern outer disc of M31, a region which includes the area
where Battinelli et al. (2003) have identified nearly one thousand
C stars.  In the outer M31 region not
previously surveyed,  we identify 361 new C star candidates, having similar
photometric properties to the known ones, and confirm the
slight decrease in the luminosity of C stars with galactocentric distances.
We show
 that the Sloan g$'$, r$'$, i$'$ filters are a viable
approach, comparable to 
(CN -- TiO), to identify C stars. We  find
that the 
(g$'$ -- r$'$) colours of cool C stars can be so red that 
prohibitively long g$'$ exposures are needed to acquire faint extragalactic
C stars. This makes the  Sloan filters a less promising approach to 
extend a C star survey to several Mpc.  
Our uniform large field survey detects the edge of M31 disk at $\sim$ 35 kpc.
The intermediate-age population, represented by C stars, extends further 
to $\sim$ 40 kpc. 
 }

\titlerunning{C stars of M31}

   \maketitle
%

\section{Introduction}
Photometric identification of intermediate-age carbon stars in nearby
galaxies has recently been done using a combination of broad and narrow-band
photometry. The (CN -- TiO) technique has been applied by, for example,
Battinelli et al. (2003), Nowotny et al. (2003) and
Harbeck et al. (2004) while the near infrared approach has been adopted
by, for example, Demers et al. (2002) and Cioni \& Habing (2005). 
The SLOAN  
Digital Sky Survey
 (SDSS) photometric
system, described by Fukugita et al. (1996), has been used to 
identify carbon stars in the Galactic halo. Indeed, Krisciunas et al. (1998)
were first to show that carbon stars can be differentiated from M stars in
the  (r$'$ -- i$'$) --- (g$'$ -- r$'$) diagram. 
Faint halo carbon stars were discovered,
using this technique, from the SDSS database by Margon et al. (2002).
The first catalogue of these halo C stars has recently been published by
Downes et al. (2004) where 251 stars were identified in a 3000 deg$^2$ area.
The authors evaluate that over 50\% of the sample constitute nearby dwarf
C stars. Obviously only a global survey like the SDSS can tackle the 
Galactic halo. Surveys of nearby galaxies can, however, be done with
a more conventional approach. 

Battinelli et al. (2003) have identified nearly one thousand C stars in
the southern outer disk of M31. Approximately 0.6 deg$^2$ were surveyed
using the CFH12K mosaic. C stars are identified from their position on
the (CN -- TiO) vs (R -- I) colour-colour diagram. The reliability of this
technique is quite secure, as it was demonstrated by Brewer et al. (1995) and
Albert et al. (2000).
The M31 sample offers the opportunity to compare the (CN -- TiO) technique
with the SDSS colour-colour approach. We therefore describe here our
new observations which include part of the
 fields already observed by us and extend the
M31 disk survey to slightly larger radii.

\section{Observations and data reduction}
The observations presented in this paper consist of one Megacam field,
centered on $\alpha$ = 00:37:02.3, $\delta$ = +39:40:50 (J2000.0), 
obtained in Service
Queue Observing mode in August 2003. The Megacam camera is
installed at the prime focus of the 3.66~m Canada-France-Hawaii
Telescope. The camera consists of a mosaic of 36 2048 $\times$ 4612 pixels CCDs
providing a field of view of nearly one deg$^2$, with a resolution
of 0.187 arcsecond per pixel.
Images were obtained through g$'$, r$'$ and i$'$ SDSS filters. The observations
were secured under non photometric and partly cloudy conditions. For this
reason, three short exposures were taken under excellent conditions and
have been used to calibrate the long exposures. It turns out, however, that
our adopted exposure times are somewhat too short to fully survey the very red
stars of M31. Table 1 presents the journal of the observations.  
The g$'$, r$'$ and i$'$ magnitudes, are calibrated with bright first
generation SDSS standards (Smith et al. 2002), as explained in the 
CFHT/Megacam website. 
   \begin{figure*}
   \centering
\includegraphics[width=10cm]{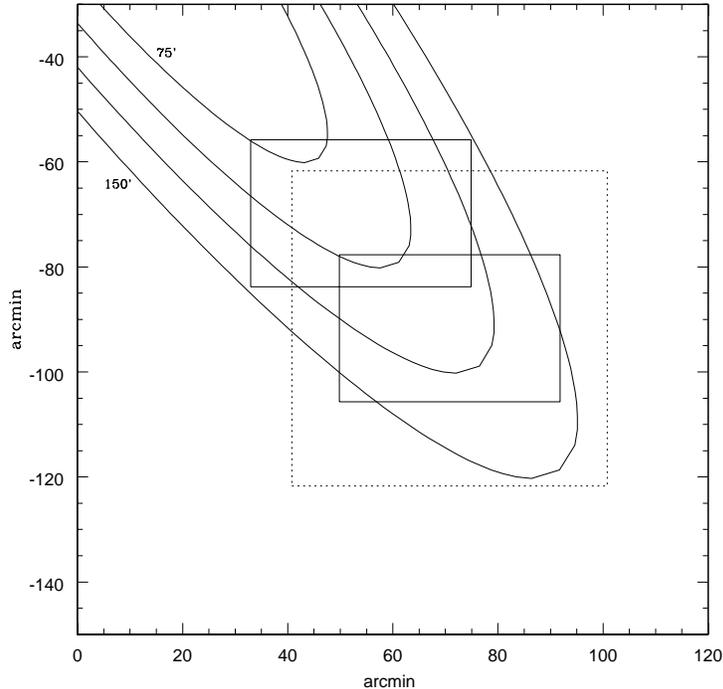}
   \caption{Schematic representation of the south west disk of M31.
Dotted square represents the Megacam field under discussion.
Solid rectangles correspond to the SW1 and SW2 fields from 
Battinelli et al. (2003). The zero point of the coordinates coincides
with the center of M31. 
               }
              \label{FigMAP}
    \end{figure*}

The data distributed by the CFHT have been detrended. This means that the
images have already been corrected with the master darks, biases, and
flats.  This pre-analysis produces 36 CCD images, of a given 
mosaic,  with the same zero point and magnitude scale. 
   \begin{table}
      \caption[]{Summary of the observations}
    $$
       \begin{array}{lllcc}
            \hline
            \noalign{\smallskip}
           {\rm  Date}&{\rm Filter}&{\rm exposure}&{\rm seeing}\ ('')& {\rm airmass}  \\
           \noalign{\smallskip}
            \hline
            \noalign{\smallskip}
2003/08/24&r'& 430\ s&1.34&1.12\\
2003/08/24&r'& 430\ s&0.93&1.12\\
2003/08/24&i'& 200\ s&0.78&1.12\\
2003/08/24&g'& 550\ s&0.97&1.11\\
2003/08/24&g'&550\ s&0.91&1.12\\
2003/08/24&g'& 550\ s&0.91&1.11\\
2003/08/24&r'& 430\ s&0.78&1.09\\
2003/08/24&g'& 550\ s&0.95&1.08\\
2003/08/24&g'& 550\ s&0.90&1.07\\
\\
2003/09/18&r'& 43\ s&0.84&1.06\\
2003/09/18&i'& 20\ s&0.84&1.06\\
2003/09/18&g'& 55\ s&0.94&1.06\\
            \noalign{\smallskip}
            \hline
         \end{array}
     $$
   \end{table}
The photometric reductions were done by Terapix, the data reduction center
dedicated to the processing of extremely large data flow. The Terapix team,
located at the Institut d'Astrophysique de Paris, matches and stacks all
images taken with the same filter and, using SExtractor (Bertin \& Arnouts
1996), provides magnitude calibrated
catalogues of objects in each of the combined images. SExtrator classifies
objects into star or galaxy but the classification scheme breaks down for
faint magnitudes. It is essentially useless for the M31 stars. A flag is
attached to each object, flag = 0 corresponds to isolated object not
affected by neighbours. As can be seen from Figure 1, the northern part
of the Megacam field is closer to the center of M31. For this reason 
a substantial stellar density gradient is observed across the field thus
numerous stars with flag $\ne$ 0 are present in the northern half.
Since the astrometric
calibration of the images has been done by the CFHT Service Observing team,
we have equatorial coordinates as well as calibrated colours and magnitude
for each object in the field.

\section{Results}
\subsection{The data}
Each calibrated $i'$, r$'$ and g$'$ file contains about 300,000 objects, 
$\sim$230,000 of them with flag = 0. We shall employ and analyse this subset. 
The number of stars, having a photometric error
smaller than a given value,  varies from one file to the
next. For example,  the $i'$, r$'$ and g$'$ files contain respectively 83,000,
121,000 and 110,000 stars with err  $<$ 0.10 mag. However, when $i'$ and r$'$
are combined, the number of stars with colour error,
$\sigma_{ri}$ $<$ 0.10, is 50,500 while
for g$'$ and r$'$, the number of stars with $\sigma_{gr}$ $<$ 0.10 is 56,700.
When the three files are combined and, following our standard criterion
only stars with $\sigma_{irg}$ =  $(\sigma_{ri}^2 + \sigma_{gr}^2)^{1/2}$ $<$ 0.125 
are retained, the number of stars drops to 37,000. We shall see later that
this small number is due to the presence of numerous faint red stars
not well observed  in the g$'$ and r$'$ filters. Finally the remaining $\sim$65,000
stars with flags = 1,2, or 3 and corresponding to objects affected by
close neighbours or/and originally  blended will not be used for the
magnitude and colour comparisons but will be needed later to cross-identify
known C stars.

\subsection{The colour-magnitude diagrams}
A one square degree field in the direction of M31 must obviously include
stars of different population. The major axis of its disk runs roughly
diagonally across the Megacam field, from 18 kpc to 33 kpc. At such 
distances the bulge population is completely negligible (Windrow et al.
2003), thus we
see disk and halo stars of M31, our Galactic contributions (halo and
disk stars) and numerous unresolved galaxies.  
We present, in Figure 2, the two colour-magnitude diagrams (CMD) 
corresponding to the whole Megacam field for stars with flag = 0 and
colour error $<$ 0.10. 
The broken lines represent the limiting magnitudes which correspond
to the magnitudes where the luminosity function drops to 50\% of its peak 
value. These are found to be: i$'$ = 22.5, r$'$ = 22.9 and g$'$ = 24.0.

The top panel shows the i$'$ magnitudes versus the (r$'$ -- i$'$) colours.
Two features are conspicuous: the bright end of the red giant branch of M31, starting
at i$'$ $\sim$ 21 and extending far to the red; and the 
vertical ridge at (r$'$ -- i$'$) $\sim$ 0.10.
This ridge corresponds to Galactic  G dwarfs,
 at the MS turnoff, seen
along the line of sight. The colour location of this ridge is
indicative of the reddening. According to Schlegel et al. (1998) the
Galactic contribution to the reddening in this direction amounts to
E(B--V) = 0.06, which translates to E(r$'$--i$'$) = 0.04.

The second CMD (lower panel of Fig. 2) has more interesting features.
Blue main sequence stars  are well separated
from the bulk of M31's stars, differential reddening within the star forming
regions must be responsible for the diffuse appearance of the main sequence.
The narrow vertical plume at (g$'$ -- r$'$)
$\sim$ 1.4 corresponds to stars of spectral type M, as the synthetic
colours from Fukugita et al. (1996) demonstrate. This plume is also
seen in the simulation of Galactic stellar objects by Fan (1999).
The vertical ridge corresponding to Galactic G dwarf is also
seen.  Finally we see, in this CMD, a small population of stars extending
at (g$'$ -- r$'$) $>$ 1.7. We identify these stars with extreme (g$'$ -- r$'$)
colours to C stars observed for the first time in the SLOAN colours
by Krisciunas et al. (1998). This C star tail is seen to curve
{\it below the the limiting magnitude of the data},
suggesting that some C stars must have been missed.

   \begin{figure*}
   \centering
\includegraphics[width=15cm]{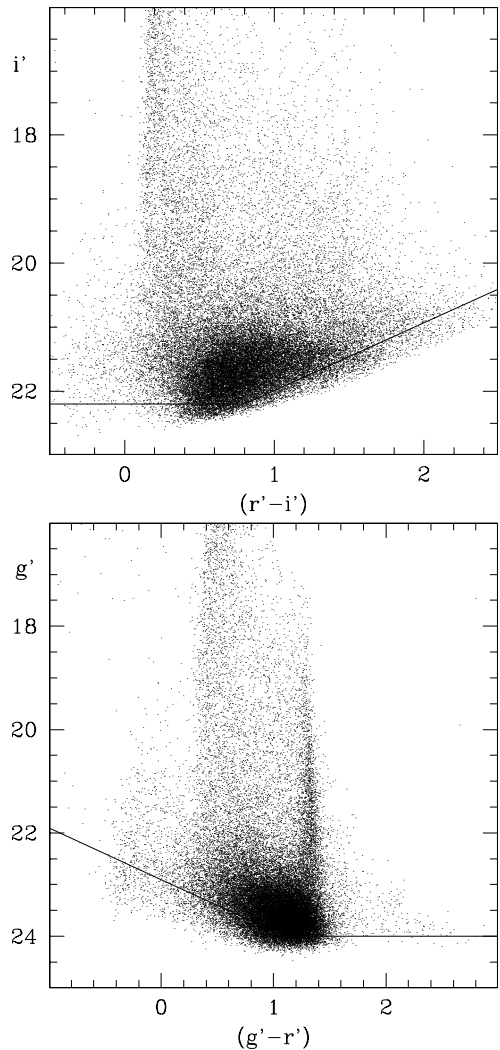}
   \caption{Colour magnitude diagrams of the stars seen in the Megacam
field. The lines correspond to the limiting magnitudes as defined in the
text. 
               }
              \label{Fig CMDs}
    \end{figure*}
\subsection{The colour-colour diagram}
The colour-colour diagram of the Megacam field is displayed in Figure 3. 
Again, only stars with flag = 0 are included,  
we plot only stars fainter than $i'$ = 17.5 to exclude some ($\sim$ 2000) of 
the foreground Galactic
stars. 
C stars are seen to the right of the diagram, with colours reaching 
(g$'$ -- r$'$) = 3. 

We note, however, a number of points with extreme 
negative (r$'$ -- i$'$) colours, seen at the bottom of the diagram.
Investigation of their spatial distribution reveals that they 
are nearly all on the West side of the 
North-South CCD borders. The spurious  $r'$  magnitudes close to the borders
are due to a $\sim$80 pixel shift of one of the exposures 
relative to the two other ones.
Thus, the instrumental magnitudes of about 200 stars near the CCD borders 
are wrong because they are not based on the right number of exposures.
It is rather difficult to delete these stars from the database because
we have lost their original x,y CCD coordinates since we are now using
equatorial coordinates. 
   \begin{figure*}
   \centering
\includegraphics[width=10cm]{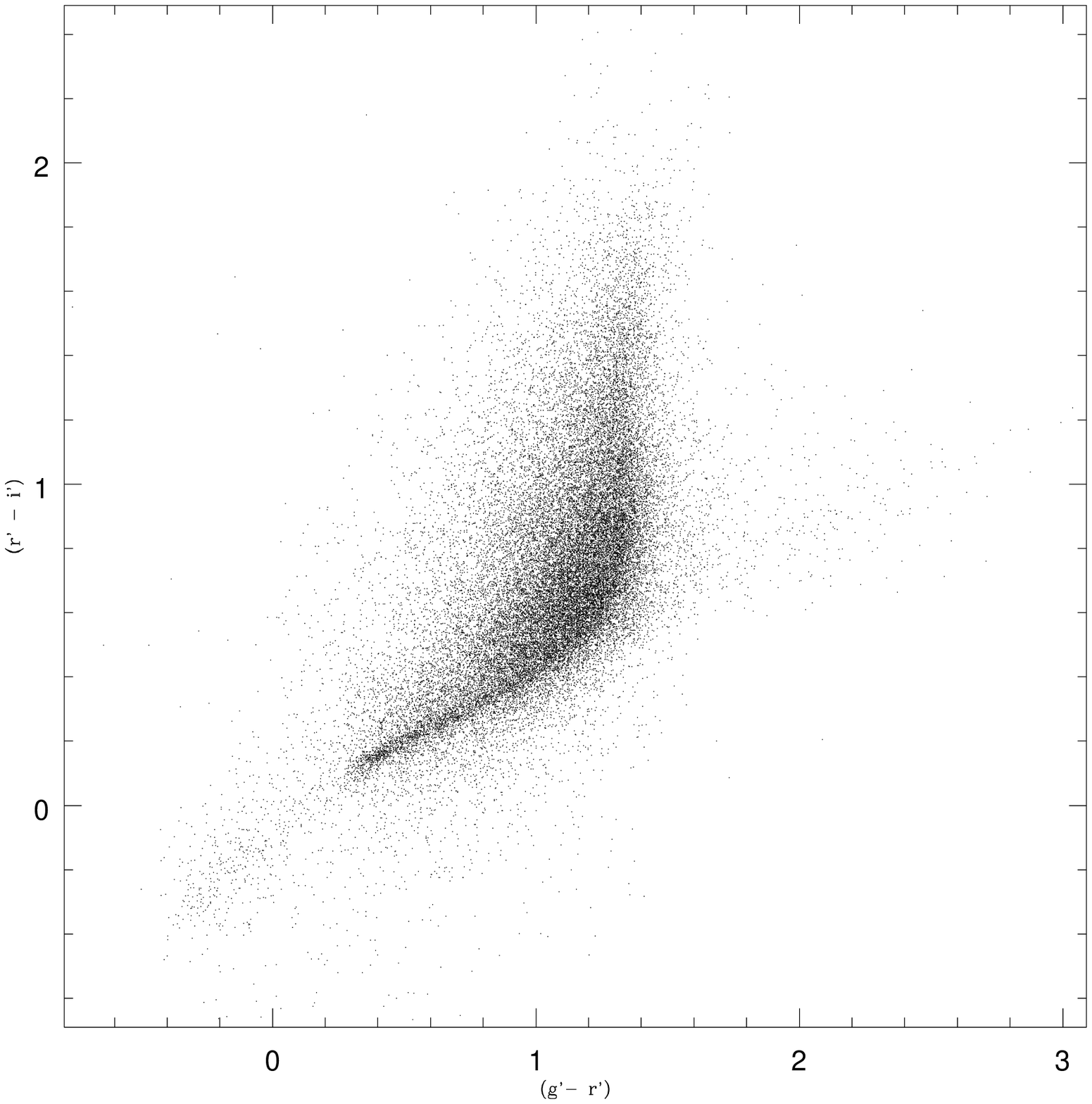}
   \caption{Colour-colour diagram of the M31 stars. 
               }
              \label{Fig CC}
    \end{figure*}

\subsection{Cross identification of M31 stars}
The goal of our study is to obtain the SDSS magnitude and colours of
the known C stars in the disk of M31 which were identified  from their 
position in the (R -- I) vs (CN -- TiO) plane. 
The (CN -- TiO) technique is designed to identify 
cool N-type C stars. In the 
(R -- I) vs (CN -- TiO) plane, the warmer and fainter C stars are mixed
with the late K or early M stars. 
For this reason the differentiation   of C stars  and M stars is limited 
to stars with (R -- I)$_0$ $>$ 0.90. Therefore,
our approach will be somewhat different from the one employed by 
Margon et al. (2002)
in their identification of faint high-latitude carbon stars (FHLCs) because
we intend to  deal exclusively with cool red C stars.

The first step in the cross identification
is to match all stars with identical equatorial coordinates. To do so, we
select two datasets: the CFH12K SW1 field consisting of 62,007 stars 
for which we have I, (R -- I) and (CN -- TiO)  
and a second set consisting of
the 37,000
stars in  the Megacam field for which we
have i$'$, (r$'$ -- i$'$) and (g$'$ -- r$'$). 
After a few iterations to minimize the $\Delta\alpha$ and $\Delta\delta$
we retain some 8000 pair of stars matched within 0.8 arcsec. The  
reason why such  relatively small number of matches is obtained can be
seen from Fig. 1, only 63\% of SW1 overlaps with 20\% of the Megacam field.
In the region common to both  fields
 there are $\sim$ 10,000 stars from
Megacam and $\sim$ 35,000 in the SW1 field.

We compare, in Figure 4,  the magnitudes and colours of these matched stars.
The i$'$ magnitudes are fainter than the I magnitudes by $\sim$ 0.25 mag.
The downward bulge, seen at the faint magnitude end, is simply due to the
natural increase of the scatter for the faintest magnitudes. The asymmetry of 
the dispersion is explained by the missing stars with faint i$'$ magnitudes.
These stars are present in the i$'$ file 
but, since they don't have
matches in the r$'$ or g$'$ files,  disappear from the i$'$r$'$g$'$ Megacam dataset.

The relationship between the magnitudes can be expressed in the following
ways, obtained from linear regressions:
$$ i' = I + (0.289\ \pm 0.010) + (0.1404\ \pm 0.0078)\times(R-I)\eqno(1)$$
or
$$ I = i' - (0.291\ \pm 0.008) - (0.1717\ \pm 0.0081)\times(r'-i').\eqno(2)$$
The comparison of the colours shows a significant scatter at the
red end. This is simply a consequence of the rather bright
r$'$ limiting  magnitude
 of our observations. The faint red stars have a lower
photometry quality.  
The colour relationship is obtained by using only 900 stars with
i$'$ $<$ 20.5. These stars have (r$'$ -- i$'$) $<$ 2.0.
$$ (r'-i') = -(0.129 \pm 0.017) + (0.857\ \pm 0.013)\times(R-I),\eqno(3)$$
this is to be compared to the synthetic colours relation
calculated by Fukugita et
al. (1996). They quote (r$'$ -- i$'$) = 0.98(R -- I) -- 0.23; for (R -- I)
$<$ 1.15 and (r$'$ -- i$'$) = 1.40(R -- I) -- 0.72; for (R -- I) $>$ 1.15.
We do not see in our data a 
break
in the colour relation up to (r$'$ -- i$'$)
$\approx$ 2.0, where the brighter stars can be seen.
For comparison, the Fukugita et al. relation is drawn on Fig. 4. 

   \begin{figure}
   \centering
\includegraphics[width=10cm]{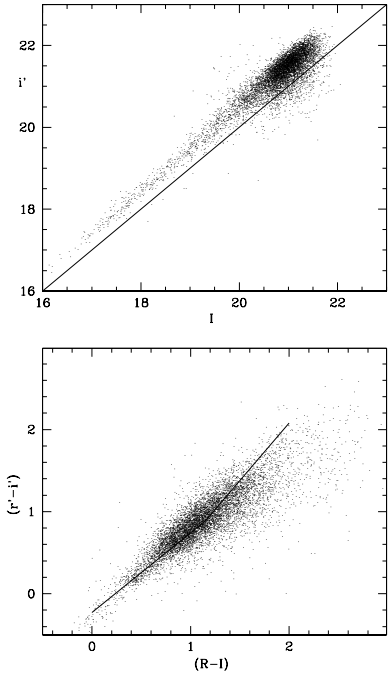}
   \caption{Comparison of the magnitudes and colours of the 8000 stars
matched between the two datasets. The top panel shows that the i$'$ magnitudes
are fainter than the I magnitudes, the identity line is drawn. 
For the bottom panel, the lines represent  the Fukugita et al. (1996)
colour relation.
               }
              \label{FigComp}
    \end{figure}
%

\subsection{Cross identification of C stars}
Of the nearly one thousand M31 C stars identified by Battinelli et al. (2003)
in fields SW1 and SW2, only 644 are in the Megacam field. 
 Cross identification, with the
same criterion described above, yields barely 129 matches, just
20\% of the C stars in the field. There are two major reasons 
for this low success
rate. The disk of M31 shows a gradient of stellar surface density. The northern
part of the Megacam field being more densely populated contains more C stars
that are in a somewhat crowded environment. Many northern C stars
were not matched.
If we do not take into account the flag assigned by SExtrator 
to each star, and accept
all stars irrespective of their flag, the number of carbon star 
matched increases to 245. This is done without relaxing the photometric error
criterion. The second explanation for the low success rate 
comes from the fact that the limiting r$'$ and g$'$ magnitudes are not
as faint as the CFH12K data, thus the redder C stars are missed. 
Figure 5 compares the (R -- I)$_0$ colour distribution of the known C stars with
those cross identified in the i$'$ and g$'$ files. Obviously the red
stars are missing in the g$'$ file and also, to a lesser extent in the 
r$'$ file, not shown here. 
%
  \begin{figure}
  \centering
\includegraphics[width=6cm]{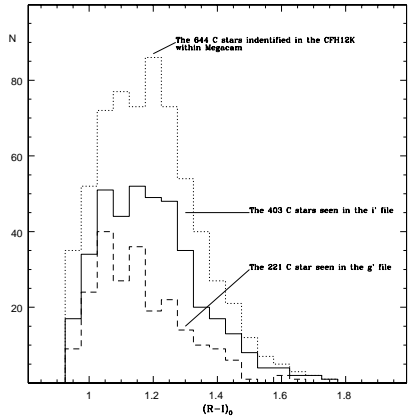}
  \caption{Colour distributions of the C stars and those cross-identified
with the Megacam data.
              }
             \label{colour distributions}
   \end{figure}

The magnitude and colour distributions of the known C stars and those
identified in our Megacam data are shown in Figure 6. The top panel
indicates the number of C stars cross identified depends 
moderately
 on the 
apparent i$'$ magnitude. This implies that the crowding is the most 
important factor.  The lower panel confirms that red
stars are missing from our Megacam sample.
  \begin{figure}
  \centering
\includegraphics[width=10cm]{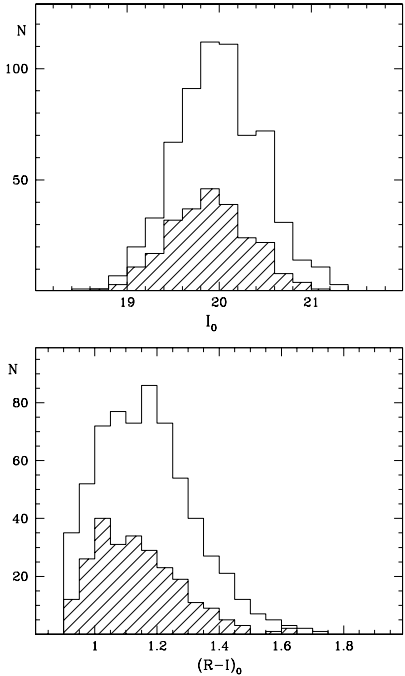}
  \caption{Magnitude and colour distributions of the  644  known  C stars are
compared to the distribution of the 245 C stars cross-identified 
with our three magnitude Megacam file (shaded histograms).
              }
             \label{Mag and colour distributions}
   \end{figure}

Figure 7 presents the SDSS colour-colour diagram of the 245 C stars 
recovered in
our Megacam data, when no flag restriction is applied. 
The three solid lines outline the acceptance limits adopted
by Margon et al. (2002) for the FHLCs. As expected, our C stars are 
redder than these limits in (r$'$ -- i$'$) as well as in (g$'$ -- r$'$).
Since our earlier adopted limit of (R -- I)$_0$ = 0.90 corresponds
to (r$'$ -- i$'$) = 0.64, very few if any C stars should have bluer 
(r$'$ -- i$'$) colours.
These Megacam observations,
 sampling the disk of M31 where numerous K giants are present, 
are certainly not ideal to identify
bluer C stars which can easily be confused with the 
bulk of the K giants. We draw, somewhat arbitrarily, the dashed lines 
corresponding to our adopted colour limits for the M31 cool C star population.
The extreme upper and lower points correspond to stars that have 
calculated magnitude and colour (given by Eqs 2 and 3) which differ
substantially from the CFH12K  values. The variability of C stars 
could be responsible or, more likely, 
spurious matches are
always possible
in a crowded field. The 129 C stars matched to the flag = 0 data are
indistinguishable in the colour-colour plane.

  \begin{figure}
  \centering
\includegraphics[width=8cm]{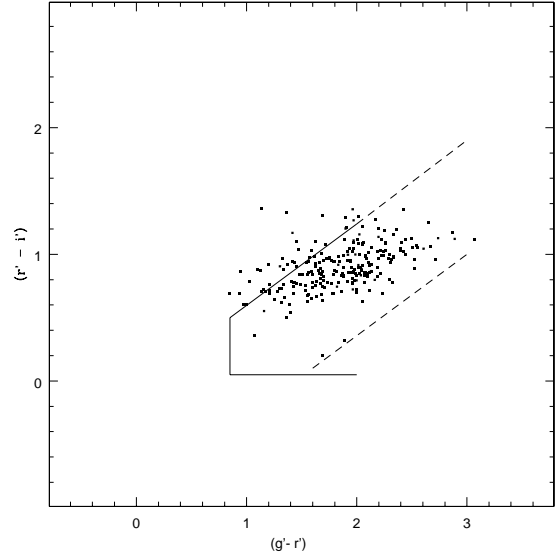}
  \caption{Colour-colour diagram of the 245 C stars, identified from their
(CN -- TiO) index. The three solid lines define the C star acceptance limits
adopted by Margon et al. (2002). The dashed lines trace our adopted
upper and lower
limits.
              }
             \label{Colour-colour plot}
   \end{figure}

\subsection{Selection criteria for cool C stars}
A close-up of Fig. 3, along with our adopt boundaries for C stars is displayed
in Figure 8. It is obvious that the blue limit adopted by Margon et al. 
(2002) is of little use to us. Indeed, the numerous K stars seen in the 
M31 disk, overwhelm the few C stars with (g$'$ -- r$'$) $\approx$ 1.5.
This, to a so large extent, that to exclude as much as possible K  and M stars
it would seem necessary  to adopt
a conservative blue limit around
 (g$'$ -- r$'$) $\approx$ 1.7.
We describe how we can better determine this colour limit.

  \begin{figure}
  \centering
\includegraphics[width=8cm]{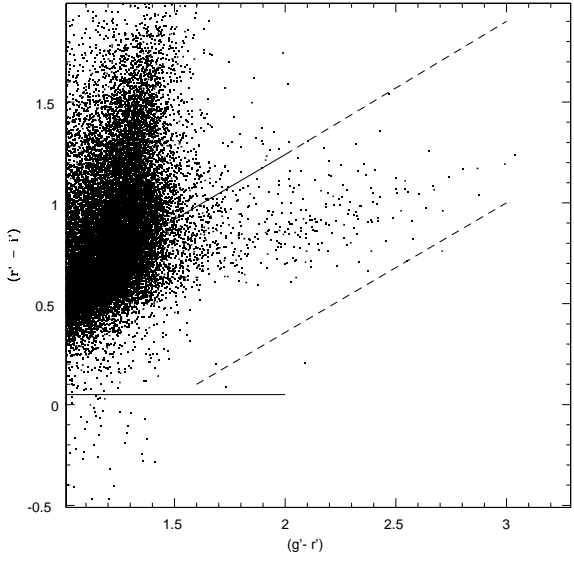}
  \caption{Close-up of the colour-colour diagram for stars with flag = 0.
The dashed lines represent the adopted boundaries.
              }
             \label{CC plot}
   \end{figure}


For the 8000 stars cross-identified we have their (R -- I), (CN -- TiO),
(g$'$ -- r$'$) and (r$'$ -- i$'$). We can then identify the C and M stars
by applying our criteria based on (R -- I) and (CN -- TiO) colours. Thus we
can calculate the number of C or M stars having a (g$'$ -- r$'$) larger
than a certain limit. 
The numbers of C 
stars (N$_C$) and  M stars (N$_M$)
 selected for different
(g$'$ -- r$'$) lower limits are displayed in Figure 9. As we shift the blue
limit to
redder colours the number of M stars drops appreciably while the number
of C stars decreases slightly. 
Figure 10 presents the variation of the ratio of the number of C stars 
to the sum (N$_C +$ N$_M$)
for various (g$'$ -- r$'$) limits. The dashed curve
is the ratio of N$_C$ relative to the original 245 C stars. 
We conclude, from this figure, that a reasonable colour limit 
is
(g$'$ -- r$'$) = 1.55. At this limit 86\% of the stars
are C stars
and we lose 20\% of the C stars which have bluer colours. 
Furthermore, as we have previously explained, the C stars detected by this
technique represent only a fraction of the total cool C star population
because our Megacam observations do not have sufficiently deep exposures.

  \begin{figure}
  \centering
\includegraphics[width=8cm]{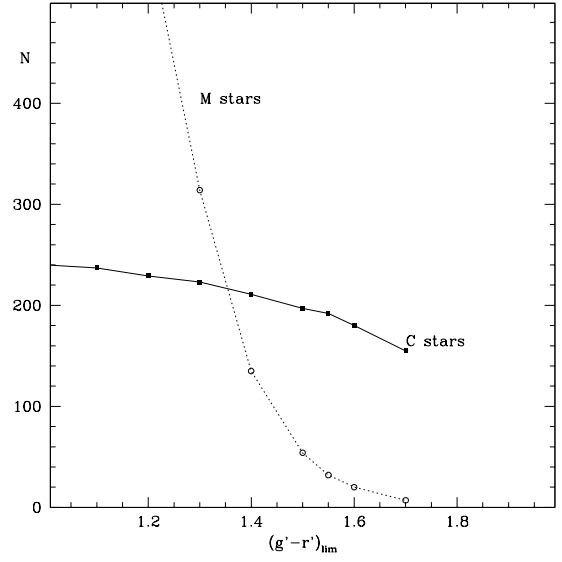}
  \caption{ The numbers of C and M stars retrieved from samples having
different colour limits show that the selection of M stars is very 
sensitive to the colour limit.
              }
             \label{C and M stars}
   \end{figure}

  \begin{figure}
  \centering
\includegraphics[width=8cm]{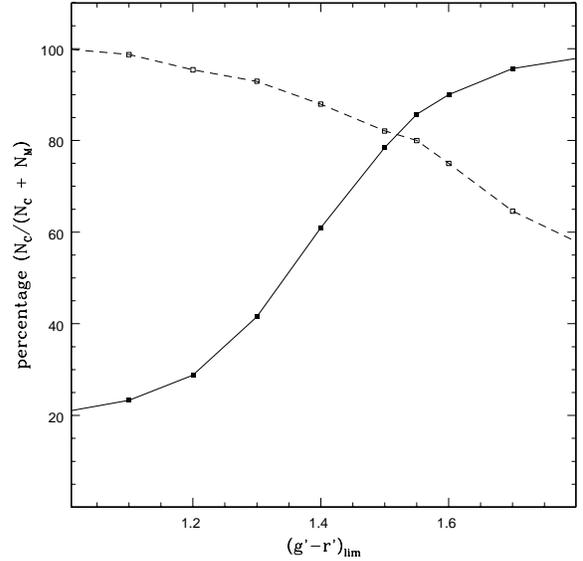}
  \caption{ The change of the ratio (N$_C$/(N$_C$ + N$_M$) for 
different colour limits. The dashed curve is the percentage of the 245
C star retained.
              }
             \label{ratios of C/M  stars}
   \end{figure}

\section{Discussion}
\subsection{Properties of C stars}
Applying the above colour limits to our Megacam data file, for stars
with flag = 0 and colour $\sigma_{irg}$ $<$ 0.125, 
yields a sample of 480 C stars 
candidates. According to Fig. 9, there should be some 70 K or M stars polluting 
this sample. Cross-identification with the known C stars in M31 results in 102 matches.
However, a cross-identification with the whole CFH12K SW1 database reveals that
17 stars are in fact M stars. These few stars are thus deleted from our sample.
The equatorial coordinates, given in degrees, the magnitude, 
the colours and their attached errors as determined by SExtractor 
of these remaining 463 stars are listed in Table 2. The $\sigma_{gr}$
being twice as large than the $\sigma_{ri}$ is explained by the 
faintness of C stars in the g$'$ band. 361 of them
are newly identified C stars located almost all at larger radial distances
than the ones previously known.

The i$'$ luminosity function of the 463 C stars candidates 
is displayed in Figure 11.
A Gaussian is fitted by eyes over the distribution. Their mean apparent 
magnitude $\langle i' \rangle$ = 20.64, with a variance of 0.31, their mean 
colours are:  
$\langle(r' - i')\rangle$ = 0.86 and $\langle(g' - r')\rangle$ = 1.86.
Using eq. 2, this mean magnitude  corresponds to
$\langle I \rangle$ = 20.20, a value to be compared with 
$\langle I_0 \rangle$ = 19.94, obtained for the M31 C star population by
Battinelli et al. (2003). Taking into account a mean extinction, of the order
of A$_I$ = 0.12, implies that we have
acquired essentially the same stellar population even though using different
colour criteria.


\begin{table*}
     \caption[C star candidates in the outer disk of M31]{$^{\mathrm{a}}$}
   $$
      \begin{array}{lcccccccc}
           \hline
           \noalign{\smallskip}
           id&RA&Dec&i'&\sigma_{r}&(r'-i')&\sigma_{r-i}&(g'-r')&\sigma_{g-r}  \\
          \noalign{\smallskip}
           \hline
           \noalign{\smallskip}
    1& 8.6219311 &39.6322746& 20.877&  0.030&  0.994&  0.043&  2.415&  0.113\\
    2& 8.6224728 &39.7282753& 20.107&  0.018&  0.850&  0.025&  1.791&  0.042\\
    3& 8.6282034 &39.8302460& 20.585&  0.026&  1.134&  0.040&  2.209&  0.097\\
    4& 8.6309576 &39.3995628& 20.726&  0.028&  0.471&  0.034&  1.571&  0.042\\
    5& 8.6383266 &39.3832932& 20.548&  0.036&  0.899&  0.048&  1.721&  0.073\\
    6& 8.6383543 &39.8645439& 21.190&  0.036&  0.950&  0.051&  2.027&  0.107\\
    7& 8.6440039 &39.7481689& 21.465&  0.043&  0.610&  0.054&  2.024&  0.093\\
    8& 8.6491823 &39.6935310& 20.527&  0.022&  0.974&  0.032&  2.636&  0.097\\
    9& 8.6501064 &40.1469650& 19.649&  0.018&  0.793&  0.023&  1.691&  0.035\\
   10& 8.6624346 &39.3817787& 20.508&  0.021&  1.049&  0.033&  2.192&  0.075\\
   11& 8.6695499 &39.4705505& 20.596&  0.023&  0.917&  0.032&  2.112&  0.067\\
   12& 8.6699352 &39.5931396& 20.262&  0.020&  0.844&  0.028&  1.918&  0.051\\
           \noalign{\smallskip}
           \hline
        \end{array}
    $$
\begin{list}{}{}
\item[$^{\mathrm{a}}$] Complete Table 2 is available in electronic form at
the CDS via anonymous ftp to cdsarc.u-strasbg.fr (130.79.128.5). 
A portion is shown here for guidance regarding its form and content.
Units of right ascensions and declination (J2000) are in degrees.

\end{list}
   \end{table*}

  \begin{figure}
  \centering
\includegraphics[width=8cm]{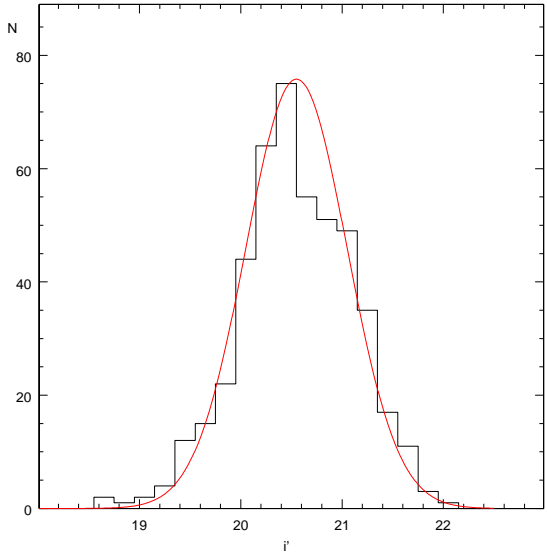}
  \caption{ Luminosity function of the 463 C stars identified by applying
our colour criterion.
              }
             \label{C LF}
   \end{figure}

\subsection{The M31 disk and its intermediate-age population}
C stars
can be used to
map the spatial distribution of the intermediate-age population.
 To do so, we calculate the surface density
per arcmin$^2$ in elliptical annuli of 10$'$ width having the shape of
the M31 apparent disk. Such ellipses are sketched in Fig. 1.
We adopt for the disk of M31 a position angle of the major axis of 37.7$^\circ$
and an ellipticity $\epsilon$~= 0.787. 
Because of the huge size of M31, 
even compared to our one deg$^2$ field, we are observing only 
sectors of annuli of quite different angular length. Even though the
radial distance along the major axis reaches only $\sim$ 150$'$, we can sample
larger radial distances on the edge of the field away from the major axis.
The top panel of Figure 12 presents the C star surface density as a function
of the radial angular distances. The bottom
panel shows the surface density of the all stars detected by
SExtractor fainter than i$'$ = 17.5 to exclude the brighter Galactic stars.
These stars are mostly M31's red giants. Their
surface density reaches a plateau at $\sim$ 170$'$
($\sim$ 35 kpc) this position should correspond to
 the edge of the stellar disk. 
If we adopt, by averaging the last 7 points,
 36.54 stars/arcmin$^2$ for the density outside of the disk,
the declining slope of the surface density corresponds to a scale length
of 4.9 $\pm$ 0.4 kpc. This is in excellent agreement with the scale length
of C stars determined by Battinelli et al. (2003) and the one derived
by Walterbos \& Kennicutt (1988) from multicolour integrated surface 
photometry. 

The density profile of 
C stars is much more irregular. The  
flattening of the profile
at short
distances is most probably due to the fact that SExtractor has difficulties
dealing with crowded fields. For C stars only stars with $\sigma_{irg}$ $<$
0.125 are selected while stars of all errors are included in the bottom
panel. 
Contrary to the bulk of stars, C stars are seen up to 
$\sim$ 190$'$, ($\sim$40 kpc) where a sharp drop is observed. The last
three points represent only
five C stars farther than 200$'$.
This confirms the identification of C stars along the major axis of the
disk by Battinelli \& Demers (2005) who found one C star at 40 kpc, well
outside of the Megacam field.

The fact that C stars are seen beyond ``the edge of the disk'' suggests
that a tenuous thick disk, containing intermediate-age stars, must be
present beyond the detectable
edge. It density contrast, relative to the halo
population, may be too low for easy detection. Since C stars are seen
behind a zero foreground they can be seen even in extremely low density
environment.

  \begin{figure}
  \centering
\includegraphics[width=8cm]{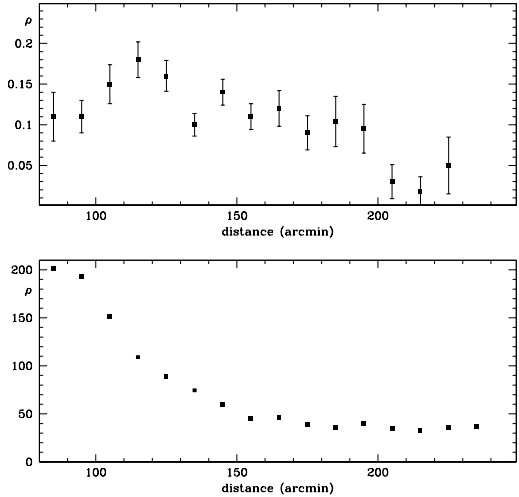}
  \caption{ The surface densities, stars per arcmin$^2$ for C stars (top)
and all stars with i$'$ $>$ 17.5 (bottom).
              }
             \label{Density profiles}
   \end{figure}

\subsection{The C/M ratio}
We have confirmed that the SDSS colours are useful to identify cool
C stars. The next step is to adopt a colour criterion applied to the
M stars, thus allowing the determination of the C/M ratio.
According the Hawley et al. (2002), M0 dwarfs have 
$\langle(r' - i')\rangle$ = 0.91 $\pm 0.24$. Unfortunately, 
because of the width and
upper colour limit of the C star zone in the colour-colour diagram
it appears quite difficult to discriminate M stars from C stars 
near such (r$'$ -- i$'$) colour. However, M3 stars have 
$\langle(r' - i')\rangle$ = 1.29 $\pm$ 0.32 a colour that cannot be
confused with C stars. Therefore, we adopt for colour limits
of the M3+ stars: (r$'$ -- i$'$) = 1.3 and (g$'$ -- r$'$) = 1.5, to
separate them slightly from C stars. We have $\sim$ 6000 such stars 
(with $\sigma_{ir}$ $<$ 0.10) in our database.
The limiting magnitude of this sample, because of the redness of the  
stars,
reaches i$'$ $\approx$ 22, which translates into  I = 21.5.

M3+ counts, in the same 10$'$ wide elliptical annuli previously adopted,
show a plateau for distances larger than 170$'$. We average these outer
counts to obtain an estimate of the foreground/background density. In our
case it corresponds to 1.06 $\pm$ 0.03 M3+ stars per arcmin$^2$.
A value we substract from the observed M3+ density. This density is to be compared
to the M3+ density of 0.8 stars/arcmin$^2$ we calculate from Durrell's et al. (2001)
observations of a remote field near M31.

Figure 13
presents the C/M3+ ratio as a function of  galactocentric distances.
As expected, for a decrease of metallicity with radial distances, the number
of C stars increases
 relative to the number of M3+ stars. Such behavior, for example,
has been
 observed in M33 by Rowe et al. (2004).

  \begin{figure}
  \centering
\includegraphics[width=8cm]{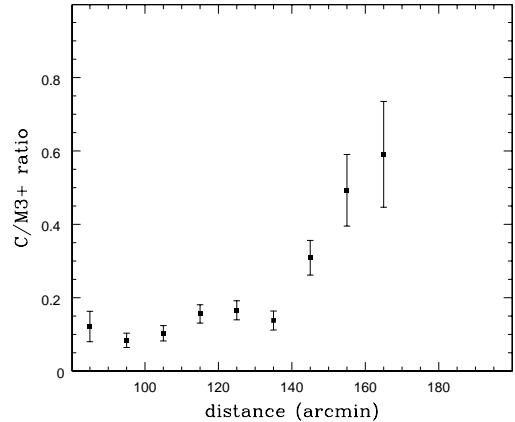}
  \caption{ The C/M3+ ratio determined for elliptical annuli of the shape
of M31's disk. 
}
             \label{C/M profile}
   \end{figure}

\subsection{Colour and magnitude trend with the galactocentric distance}
Our previous  two investigations of the outer disk of M 31 have revealed
the existence of a mild decrease of the I luminosity of C stars with increasing 
galactocentric distances (see Battinelli \& Demers, 2005). It is therefore
worth to inspect the behaviour of the newly identified C stars which extend
the previous surveys up to nearly 240$'$. Figure 14 shows the magnitude and 
colour as a function of the galactocentric distance for the newly identified
C stars along with the nearly 1000 previously know (Battinelli et al., 2003;
Battinelli \& Demers, 2005). Megacam $i'$ were converted into Kron-Cousins I 
using eq. (2).
  \begin{figure}
  \centering
\includegraphics[width=8cm]{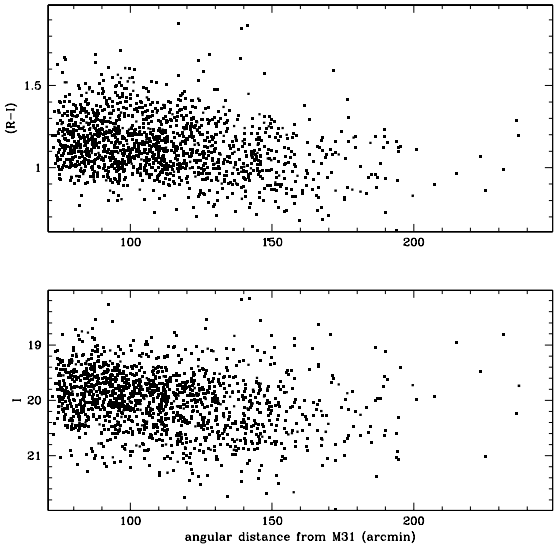}
  \caption{ Colour and Magnitude of C stars as a function of the galactocentric
distances. 
}
             \label{gradients}
   \end{figure}
A certain radial fading of the luminosity and a gradual disappearance of the
reddest C stars is evident.  In principle, the luminosity fading may be 
explained by an increase of the metallicity at large galactocentric distances. 
Indeed, it is well know that the higher the metallicity the lower is the C
star luminosity. This explanation is however unsatisfactory since both the 
observed colour gradient, shown in the top panel of Fig. 14,
 and the C/M behaviour in Fig. 13 point
definitely towards a metallicity decrease in the outer part of the Andromeda 
disk. 
Age could, however, be responsible for the observed luminosity 
fading. Indeed, theoretical models for simple stellar populations of intermediate
age (see e.g. fig. 11 in Marigo et al., 1999) suggest that
the brightest C stars 
disappear when the age of the population increases. 
A radial increase of the age of the
youngest C stars (or equivalently a decrease of the mass of the most massive 
C stars) can overwhelm the metallicity effect and explain the luminosity trend.
It is clear that, beyond this qualitative considerations, an
answer to the question is possible only through a full modeling -- which
could also account for the 
observed metallicity and density radial trends -- of the
composite stellar population in the outer disk.  

\section{Conclusion}

The SDSS filters offer an alternative method to identify C stars and
also late M stars. Contrary to the (CN -- TiO) technique, it is difficult
with the SDSS filters to isolate C stars from M0 stars. Furthermore,
because numerous C stars have quite large (g$'$ -- r$'$) colour the
exposure time to reach the desired g$'$ magnitude can be very long
relative to the i$'$ exposure. In term of telescope time both techniques
require approximately the same total exposures since the CN and TiO
exposures must be at least three to four times the I exposures.

The SDSS approach provides, however,
uncontested advantages for two aspects of the
extragalactic C stars survey. These filters are available on the new
generation of large mosaic detectors, such as Megacam. They allow the
survey of an entire nearby galaxy in a relatively short time. We have
recently followed this approach to survey 4 deg$^2$ around NGC 6822 where
nearly 900 C stars are already known in its extended halo
(Letarte et al. 2002). Optical imagers, albeit of small field size,
available on some large telescopes, such as Gemini, offer -
sometimes exclusively - SDSS filters. Because of the lack of the general
availability of the CN and TiO filter, the SDSS approach must be adopted
for any attempt to survey C stars among the neighbours of the Local Group.

\begin{acknowledgements}
This research
is funded in parts (S. D.) by the Natural Science and Engineering Council
of Canada. We are grateful to Yannick Mellier and the Terapix team to 
have so promptly accepted to measure our Megacam data.
\end{acknowledgements}

\end{document}